\documentclass{desyproc}

\usepackage[T1]{fontenc}
\usepackage{microtype}
\usepackage{fourier}
\usepackage[scaled=0.9]{berasans}

\usepackage[colorlinks=true]{hyperref}
\definecolor{CustomBlue}{RGB}{0,0,122}
\definecolor{CustomRed}{RGB}{173,42,26}
\hypersetup{linkcolor=CustomBlue, citecolor=CustomRed, urlcolor=CustomBlue}
\usepackage[sort&compress,numbers]{natbib}

\begin{document}
	
	\title{A Preview of Global Fits of Axion Models in GAMBIT}
	
	\author{{\slshape Sebastian Hoof (for the GAMBIT Collaboration)}\\[1ex]
		Department of Physics, Imperial College London, London, United Kingdom}
	
	\contribID{12}
	
	\confID{13889}  
	\desyproc{DESY-PROC-2017-XX}
	\acronym{Patras 2017} 
	\doi  
	
	\maketitle
	
	\begin{abstract}
		There are currently many experimental searches underway that try to detect QCD axions and axion-like particles. These, in addition to other constraints from astrophysics and cosmology, require a consistent and statistically rigorous analysis of the large landscape of axion models. Here, we present the rationale, methodology and preliminary results for global fits of axion models using the recently developed software framework~\textsf{GAMBIT}.
	\end{abstract}
	
	\section{Introduction}\label{sec:intro}
	The QCD axion first appeared in the context of the Peccei-Quinn~(PQ) mechanism, which was introduced to solve the Strong CP problem~\cite{Peccei:1977hh,Peccei:1977ur}. Both Weinberg~\cite{Weinberg:1977ma} and Wilczek~\cite{Wilczek:1977pj} realised that the PQ mechanism gives rise to a pseudo-scalar particle. The QCD axion also turned out to be an excellent dark matter~(DM) candidate~\cite{Preskill:1982cy,Abbott:1982af,Dine:1982ah,Turner:1985si}. There also exists a much broader class of axion models, dubbed axion-like particles~(ALPs), which are also well-motivated due to being DM candidates or appear in other contexts~\cite{Kim:1986ax,Jaeckel:2010ni}.
	
	It is important to note that there are multiple possibilities of how to extend the Standard Model to include QCD axions or ALPs. In fact, the initial models of KSVZ type~\cite{Kim:1979if,Shifman:1979if} or DFSZ type~\cite{Zhitnitsky:1980tq,Dine:1981rt} define a ``band'' of models in parameter space~\cite{hep-ph/9802220,Olive:2016xmw}. Recently, several authors categorised the original axion band in terms of phenomenologically interesting models as well as extended it~\cite{1610.07593,1611.09855,1705.05370,1709.06085}. Hence, many possible axion models have to be compared to the available data.
	
	Furthermore, there is already a wide range of experiments, observations, and theoretical arguments to constrain axion models. These efforts will grow in the coming years due to a continuing stream of new ideas and collaborations -- as, e.g., seen in various talks throughout this conference. We therefore expect that the existing, early proposals~\cite{Sikivie:1983ip,Sikivie:1985yu,VanBibber:1987rq,Anselm:1986gz,Anselm:1987vj} will be complemented by future detectors and observations.
	
	While the QCD axion has been studied extensively in the literature, theory uncertainties in its properties remain and ought to be taken into account. This also applies to the recent advances in lattice simulations of the temperature dependence of the axion mass~\cite{1606.03145,1606.07494}.
	
	Finally, any analysis of data is based on assumptions and, when combining different results, it is important to ensure their consistency. One example is the dependency of the signal predictions on the local DM abundance in axions for haloscope searches. The signal needs to be rescaled according to this quantity while, at the same time, one needs to consider the range of possible values for the local DM abundance.
	
	We argue that, in light of the points raised above, it is important to consistently combine all available results in a statistically rigorous way. In the following, we introduce \emph{global fits} as a method to, in principle, address and incorporate all of the issues mentioned in the introduction.
	
	\section{Global fits of axion models in GAMBIT}\label{sec:globalfits}
	The underlying idea of global fits is to test one or multiple models against all available data and prior information. The goals of such studies are, e.g., parameter extraction, model comparison and discovery assessment. In order to achieve these goals, one has to make use of powerful algorithms to explore the parameter space. Ultimately, identifying the models that ``best'' describe the data is also the goal of the recently published software framework \textsf{GAMBIT}~\cite{1705.07908,1705.07919,1705.07920,1705.07933,1705.07959,1705.07936}.
	
	\subsection{General overview}
	\textsf{GAMBIT}, the \textbf{G}lobal \textbf{A}nd \textbf{M}odular \textbf{B}eyond-the-Standard-Model \textbf{I}nference \textbf{T}ool, is a software framework for global fits. As the name suggests, one of \textsf{GAMBIT}'s main features is modularity. This allows for easily integrating new components such as additional models, likelihoods and sampling algorithms. The user has the freedom to only include the likelihoods deemed relevant for a problem and perform a Bayesian \emph{or} frequentist analysis. To this end, \textsf{GAMBIT} provides internal sampling routines and can also use a number of popular, external samplers~\cite{1705.07959}.
	
	Using this framework, the Collaboration published a number of physics studies regarding the MSSM~\cite{1705.07917}, scalar singlet dark matter model~\cite{1705.07931}, and GUT scale SUSY models~\cite{1705.07935}. These will be complemented by, amongst other studies, a global fit of axion models. For more information on \textsf{GAMBIT}, we encourage the reader to consider the sections of Ref.~\cite{1705.07908} relevant to their interests.
	
	\subsection{Results for axion global fits}
	In our first study of axions in \textsf{GAMBIT}, we consider the case where the PQ symmetry is broken \emph{before} the end of inflation~(and not restored afterwards). As a consequence, the entire Universe is contained within one causally-connected bubble with homogeneous initial conditions for the axion field~\cite{Preskill:1982cy}. However, the initial field value need not be of order one, such that constraints from overproduction of dark matter can be avoided~\cite{Pi:1984pv}. Since the initial field value~$\theta_\mathrm{i}$ is uniformly chosen from the interval of $-\pi$ to $\pi$~\cite{Axenides:1983hj,Linde:1984ti,Linde:1985yf}, this possibility comes at the price of a certain degree of fine-tuning. One also has to consider potentially problematic isocurvature perturbations~\cite{Axenides:1983hj,Linde:1984ti,Linde:1985yf} but, in exchange, an axion population from topological defects is diluted away by inflation.
	
	We want to include all of the most relevant constraints and experiments for this case, such as results from ALPS~\cite{1004.1313}, CAST~\cite{hep-ex/0702006,1705.02290}, various haloscope searches~\cite{DePanfilis:1987dk, Wuensch:1989sa,Hagmann:1990tj,astro-ph/9801286,Asztalos:2001tf,astro-ph/0310042,0910.5914}, Supernova 1987A~\cite{1410.3747}, H.E.S.S.~\cite{1311.3148}, various telescope searches~\cite{Bershady:1990sw,Ressell:1991zv,astro-ph/0407207,astro-ph/0611502}, HB stars/$R$-parameter~\cite{1512.08108} and the somewhat controversial WD cooling hints~\cite{1205.6180,1605.06458,1605.07668,1211.3389,1512.08108,1708.02111}. Most of these constraints have already been successfully implemented in \textsf{GAMBIT}.
	
	The structure of \textsf{GAMBIT} allows us to set up a family of axion models, including generalised QCD axion and ALP models as well as more specific models such as DFSZ or KSVZ. We consider an axion mass range from about $10^{-10}$~eV to $10$~eV. We will provide more details about the implementation of the various models in our upcoming publication. For this summary, it suffices to say that the QCD axion mass is given by
	\begin{equation}
		m_A(T) = \frac{\Lambda^2}{f_A} \left\{
		\begin{array}{@{}cl}
			1 & \mathrm{if \; } T \leq T_\mathrm{crit} \\
			\left ( \frac{T_\mathrm{crit}}{T} \right )^{\beta/2} & \mathrm{otherwise}
		\end{array} \right. , \label{eq:axionmass}
	\end{equation}
	where the energy scale~$\Lambda$, the exponent~$\beta$ and the critical temperature~$T_\mathrm{crit}$ are variable nuisance parameters, which are scanned over the range allowed by recent lattice results~\cite{1606.07494}. For ALPs, one simply sets~$\beta=0$. The axion-photon coupling~$g_{A\gamma\gamma}$ is given by,
	\begin{equation}
		g_{A\gamma\gamma} = \frac{\alpha_\mathrm{EM}}{2\pi f_A}\left(\frac{E}{N} - \widetilde{C}_{A\gamma\gamma}\right) \, ,
	\end{equation}
	with the anomaly factor~$E/N$ and the model-independent contribution~$\tilde{C}_{A\gamma\gamma}$  (we use the value obtained from di~Cortana \textit{et al.}~\cite{1511.02867}). For ALPs, on the other hand, we replace the term in brackets by a single, arbitrary coupling constant.
	
	\begin{figure}
		\centerline{
			\includegraphics[width=0.5\textwidth]{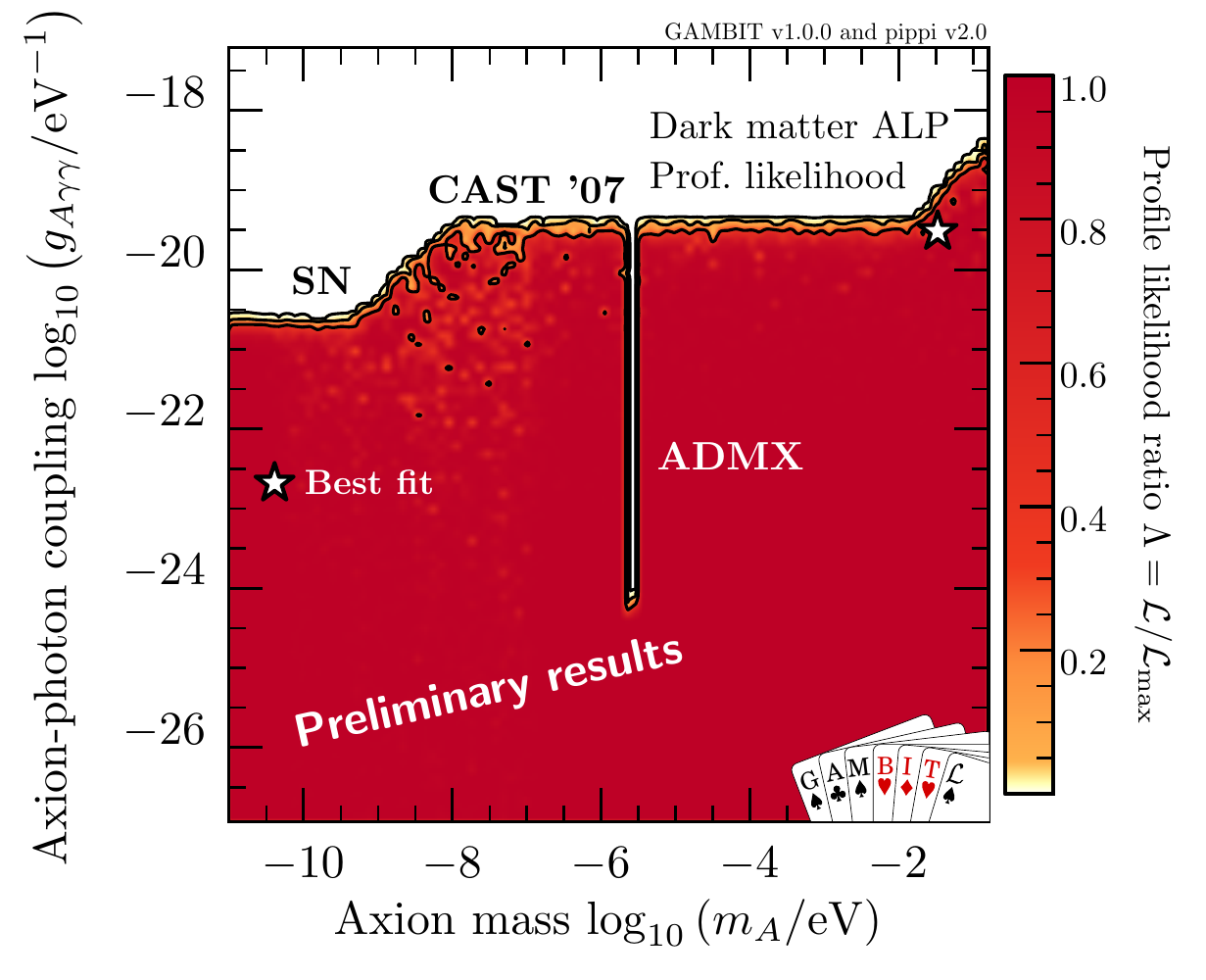}
		}
		\caption{Exclusion limits for the axion-photon coupling~(using \textsf{Diver}~\cite{1705.07959}). We used our generic ALP model and assumed axions to be all of the DM. The exclusion regions are mostly due to Supernova 1987A~(SN)~\cite{1410.3747}, ADMX~\cite{astro-ph/9801286,Asztalos:2001tf,astro-ph/0310042,0910.5914} and CAST~2007 results~\cite{hep-ex/0702006}.}\label{fig:exclusion}
	\end{figure}
	Using the already included likelihood functions, we can show that we are able to reproduce existing exclusion limits. An example for this is the $m_A$-$g_{A\gamma\gamma}$-plane as shown in Fig.~\ref{fig:exclusion} for DM made out of ALPs. For the latter, we sampled the likelihood for a general ALP model (fixing $\Lambda=10^{16}$~GeV, imposing relic density within uncertainties and sampling over the local halo density). The profile likelihood therefore follows closely the na\"{i}vely expected exclusion results from, e.g., overplotting various exclusion curves~(such as in Ref.~\cite{Olive:2016xmw}).
	
	\begin{figure}
		\centerline{
			\includegraphics[width=0.49\textwidth]{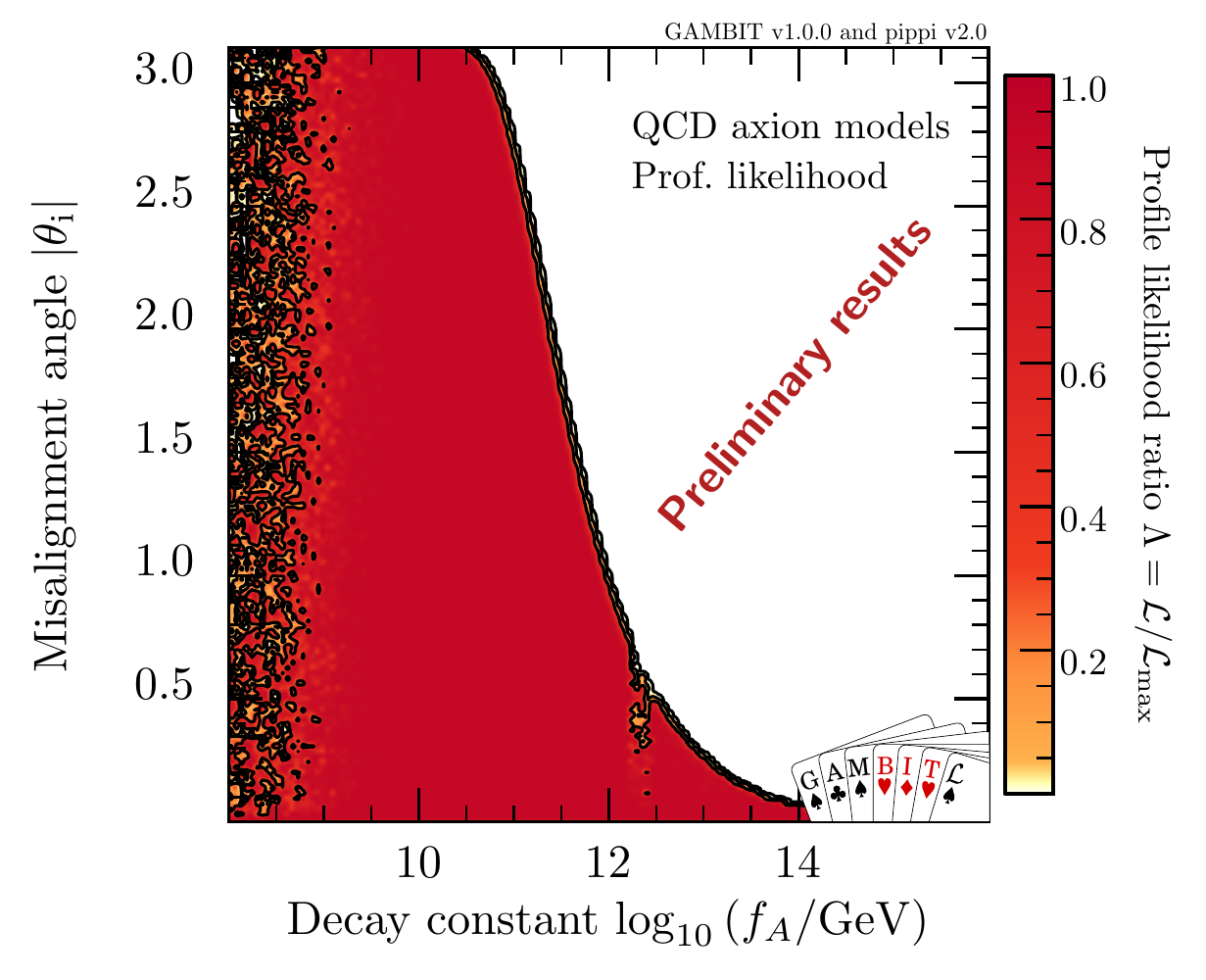}\hfil
			\includegraphics[width=0.49\textwidth]{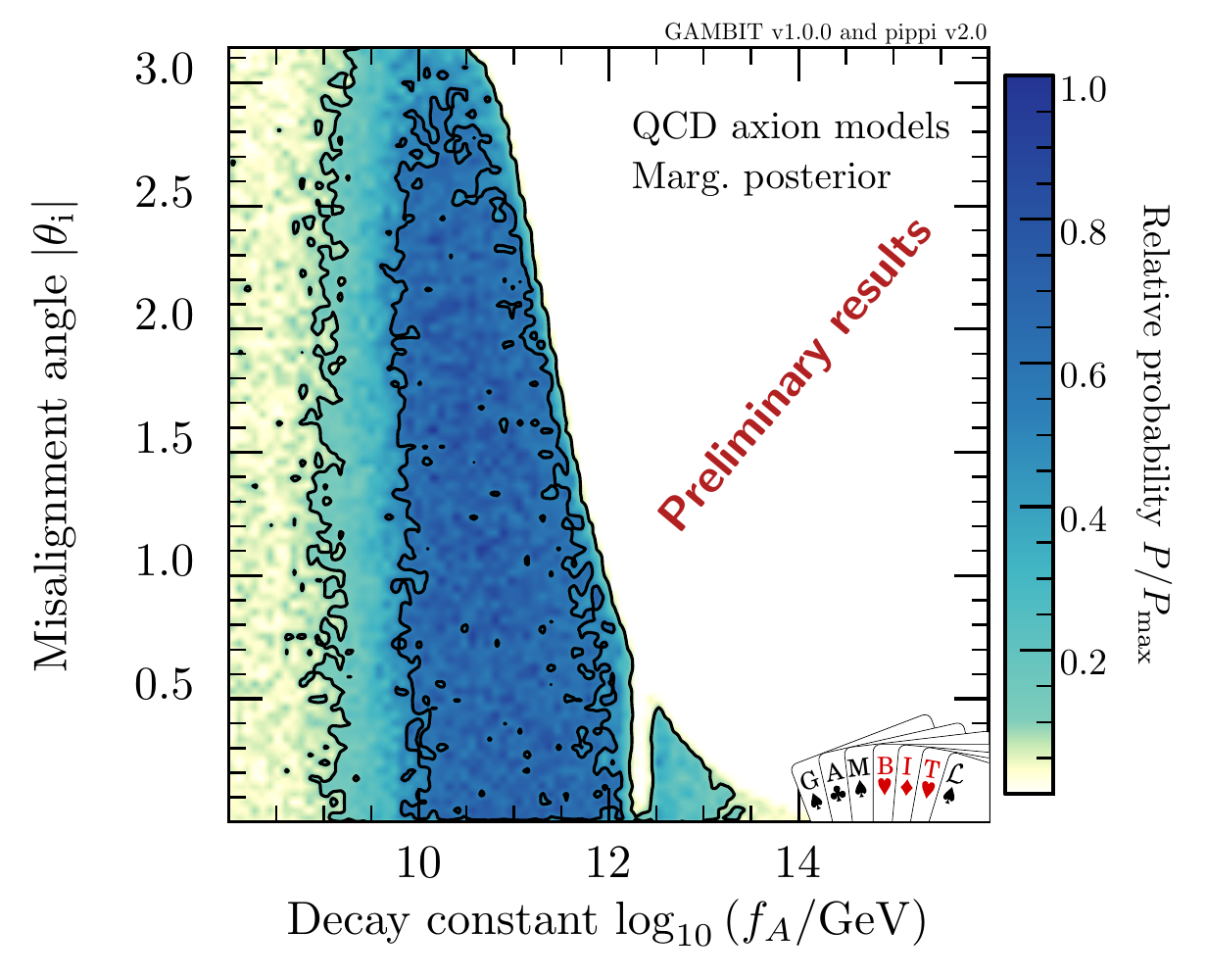}
		}
		\caption{Frequentist~(\textit{left}, using \textsf{Diver}~\cite{1705.07959}) vs Bayesian~(\textit{right}, using \textsf{MultiNest}~\cite{0704.3704,0809.3437,1306.2144}) inference. We sample~$\theta_\mathrm{i}$ uniformly in the interval of $-3.14$ to $3.14$ and only impose an upper limit on axions being the DM using Planck results~\cite{1502.01589}.}\label{fig:thetai}
	\end{figure}
	As pointed out earlier, we rescale the signal expectation for, e.g., the ADMX~experiment and are therefore sensitive to potential preferences regarding values of parameters such as~$\theta_\mathrm{i}$. Even though there is no \textit{a~priori} argument for why certain values of~$\theta_\mathrm{i}$ should be preferred over others, this prior information will still influence a Bayesian analysis. In fact, in the pre-inflationary PQ symmetry breaking scenario, $\theta_\mathrm{i}$~is uniformly distributed in the interval from $-\pi$ to $\pi$. We therefore have a physically-motivated prior with ``natural'' values of~$\mathcal{O}(1)$. The Bayesian analysis automatically captures this well-known fine-tuning argument, and the limits from ADMX persist in the right panel of Fig.~\ref{fig:thetai}, while the profile likelihood should be ignorant of that fact. However, the completeness of our sampling depends on the choice of priors. Better sampling can be achieved by combining the results of various scans (at least for the frequentist approach) or optimising our sampler settings.
	
	Improving the sampling, including the remaining likelihood functions and using the advantages of both frequentist and Bayesian methods are the remaining steps for concluding this first study. In our final analysis we will also compare with so-called ``cooling hints" for the existence of axions~\cite{1205.6180,1605.06458,1605.07668,1211.3389,1512.08108,1708.02111}.
	
	\section{Outlook}
	In order to consistently combine the increasing amount of information on axions, we propose to use the global fitting framework \textsf{GAMBIT}. This will allow us to compare different axion models in our upcoming study on the pre-inflationary PQ breaking case. The other case, where the PQ symmetry breaks after inflation, is left for future work. Furthermore, \textsf{GAMBIT} offers the opportunity to combine axions with other beyond-the-Standard-Model physics, such as in the recently proposed SMASH model~\cite{1610.01639}.
	
	\section*{Acknowledgments}
	SH would like to thank all the contact persons from the experimental collaborations and individual authors of Refs~\cite{1004.1313,hep-ex/0702006,1705.02290,DePanfilis:1987dk, Wuensch:1989sa,Hagmann:1990tj,astro-ph/9801286,Asztalos:2001tf,astro-ph/0310042,0910.5914,1410.3747,1311.3148,Bershady:1990sw,Ressell:1991zv,astro-ph/0407207,astro-ph/0611502,1512.08108,1205.6180,1605.06458,1605.07668,1211.3389,1708.02111} for helpful discussions and their data as well as the members of the GAMBIT collaboration for helpful discussions and support. SH is funded by the Imperial College President's PhD Scholarship scheme.
	
	
\renewcommand\baselinestretch{0.5}\selectfont
\footnotesize

\bibliographystyle{apsrev_mod}
\bibliography{biblio}
	
	
\end{document}